\documentclass[twocolumn,jcp,superscriptaddress]{revtex4-1}

\usepackage{graphicx,float}
\usepackage{latexsym}
\usepackage{epstopdf}
\usepackage{amssymb,amsmath}
\usepackage{color}
\usepackage{multirow}
\usepackage{hyperref}
\usepackage{graphicx}
\makeatletter
\let\Hy@linktoc\Hy@linktoc@none
\makeatother

\hypersetup{colorlinks=true,linkcolor=blue,citecolor=red}
\usepackage{verbatim}
\usepackage{bm}

\newcommand*\Laplace{\mathop{}\!\mathbin\bigtriangleup}
\begin{document}

\title{Scaling characteristics of fractional diffusion processes in the presence of power-law distributed random noise}
\author{Mohsen Ghasemi Nezhadhaghighi}
\email{ghaseminejad@shirazu.ac.ir}
\affiliation{Department of Physics, Shiraz University, Shiraz  71454, Iran}

\date{\today}

\begin{abstract}
We present results of the numerical simulations	and the scaling characteristics of one-dimensional random fluctuations with heavy tailed probability distribution functions. Assuming that the distribution function of the random fluctuations obeys L\'evy statistics with
a power-law scaling exponent, we investigate the fractional diffusion equation in the presence of $\mu$-stable L\'evy noise. We study the scaling properties of the global width and two point correlation functions, we then compare the analytical and numerical results for the growth exponent $\beta$ and the roughness exponent $\alpha$. ‌We also investigate the fractional Fokker-Planck equation for heavy-tailed random fluctuations. We show that the fractional diffusion processes in the presence of $\mu$-stable L\'evy noise display special scaling properties in the probability distribution function (PDF). Finally, we study numerically the scaling properties of the heavy-tailed random fluctuations by using the diffusion entropy analysis. This method is based on the evaluation of the Shannon entropy of the PDF generated by the random fluctuations, rather than on the measurement of the global width of the process. We apply the diffusion entropy analysis to extract the growth exponent $\beta$ and to confirm the validity of our numerical analysis. The proposed fractional langevin equation can be used for modeling, analysis and characterization of experimental data, such as solar flare fluctuations, turbulent heat flow and etc.

\end{abstract}

\maketitle
\section{Introduction}

In recent years, the study of systems displaying
heavy-tailed distributions has attracted considerable attention including physical, chemical, biological, economical and geological systems \cite{sornette,Jacobs,Klages}. Examples include, solar flare fluctuations \cite{Scafetta}, the statistics of the turbulent fields \cite{sorriso,biskamp} and turbulent heat flow \cite{kadanoff}, the global velocity of imbibition fronts \cite{planet2009}, density of a vibrated column of granular material \cite{Nowak}, particle velocity fluctuations in granular system \cite{Radjai}, front propagation in porous media \cite{Hansen2007,Asikainen2002}, propagation of cracks in solids \cite{alava2006statistical}, crackling noise \cite{sethna2001crackling,bonamy2008crackling} and  
the motion of geological faults \cite{fisher1997statistics,kawamura2012statistical}, and many others. 

The ability to summarize observations using a unified statistical model takes the great deal of interest of Physicists. One strategy for handling the unusual properties of the phenomenon addressed here is relying on the particle based models such as fractional L\'evy motion, which includes power-law statistics and long-range correlations leading to non-Gaussian and non-local anomalous behaviors \cite{bovet2015r}. In particular, fractional L\'evy motion has been used to study suprathermal ions, created by fusion reaction \cite{bovet2014}, and turbulent plasmas \cite{bovet2015}, \textit{in vivo} diffusion magnetic resonance imaging data \cite{fan2015}, solar power influx into the magnetosphere \cite{Schumann2012}, network traffic modeling \cite{Laskin2001}, solar wind fluctuations \cite{Watkin2005} and spatial variability in sedimentary rock formations \cite{Painter1994}.

We are particularly interested to understand these problems based on the investigation of the fluctuation dynamics on a macroscopic scale, which leading finally to random field associated with linear (or non-linear) stochastic partial differential equations. The propagation dynamics of random fluctuations, can be obtained by diffusion phenomena in the presence of random forcing functions. Theoretical modeling of diffusion processes in the presence of random fluctuations have been started with the famous work by Edwards and Wilkinson (EW) \cite{edwards-wilkinson}. The EW model has found interesting applications in a wide range of non-equilibrium statistical physics, \textit{i.e.} the interface dynamics \cite{surface2,nelsonbook}. In the EW model, the scalar random field $h({x},t)$ evolving via the Langevin equation,

\begin{eqnarray}\label{edwards wilkinson}
\frac{\partial h({x},t)}{\partial t} = D  \nabla^2 h({x},t) + \eta({x},t),
\end{eqnarray}
where $D$ is a constant related to the dynamical surface tension or diffusion, and $\eta({x},t)$ is assumed to be a Gaussian white noise which satisfies $\langle \eta({x},t) \rangle = 0$  and
 $\langle \eta({x},t)\eta({x}^\prime,t^\prime) \rangle = 2k_BT \delta({x}-{x}^\prime)\delta(t-t^\prime)$. 
Since the probability distribution function of the solution to Eq. (\ref{edwards wilkinson}), obeys
a Gaussian law, it would be interesting to introduce a deformed version of EW model to investigate random fluctuations with heavy-tailed distributions.

A modified version of Eq. (\ref{edwards wilkinson}) was introduced by Lam and Sander \cite{lam1993exact}. They suggested that one might describe the dynamics of the random fluctuations by a linear stochastic partial differential equation (\ref{edwards wilkinson}) in the presence of heavy-tailed noise. They answered how the statistics of height fluctuations, is affected by the presence of random noise with power-law distribution. 

The fractional langevin equation governing the dynamics of non-local random fluctuations with heavy tailed distributions, recently proposed in Ref. \cite{nezhadhaghighi2017}. According
to the scaling analysis approach, we have investigated the scaling characteristics of the fractional diffusion processes in the presence of $\mu$-stable L\'evy noise.

In this work, our aim is to consider the same model in one spatial dimension and, based on the numerical simulations, to investigate the scaling properties of the heavy-tailed random fluctuations. Starting with Langevin equation (\ref{edwards wilkinson}) to describe the dynamics of Gaussian fluctuating interfaces, we systematically change the local diffusion operator by the non-local fractional operator. We also added heavy-tailed random processes to obtain a description applicable on the study of the correlated random fluctuations with power-law probability density functions. To this end, in Sec. \ref{section 2}, we introduce our model and by means of scaling arguments we present the statistical properties of the model that is characterized by the $q$th order global width and $q$th order two point function. The existence of power-law behaviors, plus the assumption of the self-affine character of the heavy-tailed correlated random fluctuations allows us to obtain the scaling exponents of the process \textit{i.e.} growth exponent and local (global) roughness exponent.
Particular emphasis is put on the scaling characteristics of the probability distribution function (PDF). Using diffusion entropy analysis we show how the growth scaling exponent of the random fluctuation can be described by PDF dynamics. Sec. \ref{section 3}, is devoted to the analysis of the fractional Fokker-Planck equation in order to drive the power-law scaling properties of the distributions. In Sec. \ref{section 4}, finite-difference
numerical scheme is used to integrate stochastic fractional differential equation. In this section we summarize our numerical results.   In Sec. \ref{section 5}, we summarize the main findings and conclusions.

\section{Dynamics of heavy-tailed random fluctuations}\label{section 2}

In Eq. (\ref{edwards wilkinson}), the term $\nabla^2 h({x},t)$ is proportional to the diffusion of the random fluctuation which is a local operator. In the general case we are interested to replace ordinary derivative with fractional one that is non-local fractional Laplacian. For this purpose, we define the fractional Langevin equation in the presence of heavy-tailed noise as,  

\begin{eqnarray}\label{fractional edwards wilkinson}
\frac{\partial h({x},t)}{\partial t} = D_z   \frac{\partial ^z}{\partial |{x}|^z}h({x},t) + \eta_{\mu}({x},t),
\end{eqnarray}
so that for $z=2$ and $\mu=2$ we recover the standard EW model Eq. (\ref{edwards wilkinson}) (see Ref. \cite{nezhadhaghighi2017}). In the above equation, the Laplacian operator has been replaced by $\frac{\partial ^z}{\partial |{x}|^z}$, which is the fractional space derivative of order $z$. 
The fractional derivative is defined via its Fourier transform
\begin{eqnarray}
\mathcal{F}\left\lbrace \frac{\partial ^z}{\partial |{x}|^z} h({x},t) \right\rbrace \equiv -|\mathbf{q}|^z h(\mathbf{q},t),
\end{eqnarray}
where, $h(\mathbf{q},t)$ is Fourier transform of the random field $h({x},t)$. It is also possible to rewrite the fractional derivative in terms of the standard Laplacian $\Laplace$ as $\frac{\partial ^z}{\partial |{x}|^z}\equiv -(\Laplace)^{z/2}$ \cite{samko,kilbas}. The $\mu$-stable L\'evy noise $\eta_{\mu}({x},t)$ is uncorrelated noise with power-law distribution function and $\langle \eta_{\mu}({x},t) \rangle = 0$. The parameter $\mu >0$ characterizes the asymptotic power-law behavior of the stable distribution
\begin{eqnarray}\label{power law PDF}
p(x) \sim \frac{1}{|x|^{1+\mu}}.
\end{eqnarray}
We should note that the corresponding L\'evy white noise with $0 < \mu < 2$, has infinite variance and the higher cumulants \cite{Klages, stochastic book,gardiner}.

Here we are focusing on the scaling characteristics of Eq. (\ref{fractional edwards wilkinson}) in one-dimension. It is easy to show that the solution to Eq. (\ref{fractional edwards wilkinson}) is a self-affine random process. The scale transformation of space by factor $\lambda$ and of time by a factor $\lambda^\nu$ ($\nu$ is the dynamical scaling exponent), re-scales the random process $h(x,t)$ by factor $\lambda^\alpha$ as follows,
\begin{eqnarray}
h(\lambda x,\lambda^\nu t) \equiv \lambda^{\alpha} h(x,t),
\end{eqnarray}
where $\alpha$ is called the \textit{roughness} (\textit{Hurst}) exponent. Therefore, the scale transformations of equation of motion for the random fields with heavy-tailed PDF, give us
\begin{eqnarray}\label{scale transform fractional edwards wilkinson}
\lambda^{\alpha - \nu}\frac{\partial h(x,t)}{\partial t} = \lambda^{\alpha - z} D_z   \frac{\partial ^z}{\partial |x|^z}h(x,t) + \lambda^ {\gamma} \eta_{\mu}(x,t), 
\end{eqnarray}
where $\gamma = (1+\nu)(1/\mu -1)$ (see Appendix A). The solution to Eq. (\ref{fractional edwards wilkinson}), is scale invariant. 
Therefore, the values of the dynamical exponent $\nu$ as well as of the roughness exponent $\alpha$ depend on the fractional order $z$ and the asymptotic scaling power of the stable distribution $\mu$:

\begin{eqnarray}\label{scaling exponent for fEW}
\nu = z \textrm{ and } \alpha = \frac{z+1}{\mu}-1.
\end{eqnarray}

Consider a random process $\eta_{\mu} (x,t)$ in Eq. (\ref{fractional edwards wilkinson}) that is distributed with the probability density function Eq. (\ref{power law PDF}). It implies that the variance, $\langle |\eta|^2 \rangle$, is infinite. Most importantly, fractional moments $\langle |\eta|^q \rangle$ for all $q<\mu$, become finite. From this viewpoint it is evident that, the random fluctuations governed by fractional diffusion processes in the presence of heavy-tailed noise may be characterized by the
$q$th order moments. In order to quantitatively characterize the fluctuation dynamics, let us first consider the  $q$th order \textit{global} width of the random field $h(x,t)$ defined by
\begin{eqnarray}\label{roughness measure}
w_q(t,L) \equiv \left\langle \overline{|h(x,t)-\overline{h(x,t)}|^q} \right\rangle ^{1/q},
\end{eqnarray}	
where the overline denotes a spatial average over the system of size $L$, and bracket denote ensemble averaging.
For scale invariant random fluctuations, the global width satisfies the Family-Viscek scaling ansatz \cite{family85}
and have asymptotic behavior given by
\begin{eqnarray}\label{global roughness scaling}
w_q(t,L) \sim 
\begin{cases}
    t^{\beta}       & \quad t\ll t_\times \\
    L^{\alpha_g}  & \quad t \gg t_\times\\
  \end{cases},
\end{eqnarray}
where $t_\times = L^\nu$ is the characteristic time and $\beta$ is called the growth exponent. The exponent $\alpha_g$ defines the so-called global roughness exponent. Self-affinity occurs when the  $q$th order global width  scales with a single $q$-independent exponent $\beta$. The scaling exponents $\alpha_g$, $\beta$ and $\nu$ are connected through the ratio $\beta = \alpha_g/\nu$, because of the self-affine scaling properties of the model under study \cite{surface2}.

To  study  the local properties  of  the  fluctuating random field $h(x,t)$, the $q$th order two-point correlation function is defined as
\begin{eqnarray}\label{local roughness scaling}
C_q(r,t) \equiv \langle \left[ h(x,t)) - h(x^\prime,0)\right]^{q} \rangle ^{1/q} \sim r^{\alpha_l}f(r t^{-1/\nu}),
\end{eqnarray}
where the scaling function $f(u)$ behaves like
\begin{eqnarray}
f(u) =
  \begin{cases}
    u^{-\alpha_l}       & \quad u\gg 1 \\
    \text{const.}  & \quad u \ll 1\\
  \end{cases}.
\end{eqnarray} 
where $r = |{x}-{x}^\prime|$.  In the saturated regime ($t\gg L^z$), one expects $C_q$ to scale as $C_q(r)\sim r^{\alpha_l}$. The exponent $\alpha_q$'s define local roughness exponent.  This exponent is another measure to distinguish between self-affine and multiscaling properties  of the random fluctuations. For self-affine random fluctuations there is only one roughness exponent, and thus $\alpha_g=\alpha_l=\alpha$ for all $q$s \cite{surface2}. 

The scale-invariance property in the fractional diffusion
processes in the presence of $\mu$-stable L\'evy noise can be described mathematically
with the probability distribution function $\mathcal{P} [ h(x,t)]$. It is the probability that the fluctuating
non-equilibrium field in the position $x$ and time $t$ can take a specific value $h(x,t)$. Therefore, in order to investigate more precisely the scaling properties of $q$th order global width (\ref{global roughness scaling}) and $q$th order two-point correlation function (\ref{local roughness scaling}), it would be necessary to analyze and quantify the PDF of the process.  

The Lagevin equation (\ref{fractional edwards wilkinson}) is invariant under spatial translation $x\rightarrow x+x_0$ and scale transformations $x\rightarrow \lambda x$ and $t\rightarrow \lambda^z$. In
spite of these symmetries, we expect the probability distribution function in this case to follow
\begin{eqnarray}\label{scaling of prob h}
\mathcal{P} [ h(x,t)] \equiv \mathcal{P}(y=h|_{x=x_0},t,L) = \frac{1}{t^\beta L^\alpha }\mathcal{F}\left( \frac{y}{t^\beta},\frac{y}{L^\alpha} \right),
\end{eqnarray}
where in the early time, $t \ll L^z$, it takes the form 
\begin{eqnarray}\label{scaling of prob h: early time}
\lim_{t\to 0} \mathcal{P}(y,t)\equiv  \mathcal{P}_{er}(y,t) = \frac{1}{t^\beta }\mathcal{G}_1\left( \frac{y}{t^\beta} \right),
\end{eqnarray}
and in long-time limit, $t \gg L^z$, Eq. (\ref{scaling of prob h}) has the form
\begin{eqnarray}
\lim_{t\to \infty} \mathcal{P}(y,t,L)\equiv  \mathcal{P}_{st}(y,L) = \frac{1}{L^\alpha }\mathcal{G}_2\left( \frac{y}{L^\alpha} \right),
\end{eqnarray}   
where $\mathcal{G}_1$ and $\mathcal{G}_2$ are universal functions. In the next section we carry out the asymptotic behavior $\mathcal{G}_n(x) \sim 1/x^{1+\mu}$. 

According to Eq. (\ref{global roughness scaling}), the $q$th order deviation of height from its mean behaves as:
\begin{eqnarray}
w^q   \equiv \int dy |y-\overline{y}|^q \mathcal{P}(y,t,L).
\end{eqnarray}
By taking into account that the fractional diffusion process (\ref{fractional edwards wilkinson}) is invariant under  $h\rightarrow -h$ (up/dawn symmetry), we immediately obtain $\mathcal{P}(y,t,L)=\mathcal{P}(-y,t,L)$. The symmetry in the PDF, corresponds to $\bar{h}=0$. To get the scaling form of Eq. (\ref{global roughness scaling}), we have
\begin{eqnarray}\label{global roughness scaling by pdf}
w^q(t,L) \equiv 2\int_0^\infty dy y^q \mathcal{P}(y,t,L)= 
\begin{cases}
    I_1 t^{q\beta}   & \quad t\ll L^z \\
    I_2 L^{q\alpha}   & \quad t \gg L^z\\
  \end{cases},
\end{eqnarray}
where $I_n=2\int_0^\infty  u^q \mathcal{G}_{n}(u) du$. 

To get more complete description of the non-Gaussian probability distribution function of the process (\ref{fractional edwards wilkinson}), it is instructive to calculate the diffusion entropy. This method of analysis is
based on the Shannon entropy of the diffusion process \cite{information entropy}. The Shannon entropy for the random field $h(x,t)$ can be defined as,
\begin{eqnarray}\label{shannon entropy}
S(t) = - \int \mathcal{P}[h] \log \mathcal{P}[h] \mathcal{D}[h].
\end{eqnarray}
Let us suppose that $\mathcal{P}[h]$ fits the scaling condition of Eqs. (\ref{scaling of prob h}) and (\ref{scaling of prob h: early time}). To show how the entropy analysis works, let us plug Eq. (\ref{scaling of prob h: early time}) into Eq. (\ref{shannon entropy}). After a simple algebra, we get:

\begin{eqnarray}
S(t) = - \int \mathcal{G}_1(u) \log \mathcal{G}_1(u) du + \beta \log t
\end{eqnarray}
where the scaling exponent $\beta$ is the growth exponent. As a consequence, when the PDF has the scaling form of Eq. (\ref{scaling of prob h: early time}), the diffusion entropy grows log-linearly with time,
$S(t) =\beta \log t + \mathrm{const}$. In addition the entropy, $S(t) = \mathrm{const}$, for the long time limit ($t\gg L^z$) and it depends on the system size.

\section{Fractional Fokker-Planck equation for heavy-tailed random fluctuations}\label{section 3}

The stochastic part $\eta_\mu(x,t)$ denotes a $\delta$-correlated, L\'evy white noise which is composed of independent identical random variables that are distributed according to the stable density with the characteristic index $\mu$ ($0<\mu<2$); \textit{i.e.} $P_{\mu}(x)$. The characteristic function $\phi(k)$ of a $\mu$-stable random variable $x$ can be represented by 
\begin{eqnarray}\label{characteristic function levy pdf}
\phi(k) \equiv \int_{-\infty}^{\infty} e^{-ikx}P_{\mu}(x) dx = e^{-c ^{\mu} |k|^{\mu}}~,
\end{eqnarray}
where $c$ is the scale factor \cite{Klages}.
From above definition it is obvious that the parameter choice $\mu=2$ yields the usual, Gaussian probability distribution $P_2(x) = (4\pi c^2)^{-1/2} \exp(x^2/4c^2)$ with the variance $\langle |x|^2 \rangle=2c^2$. 
The closed form for the symmetric L\'evy distribution when $\mu = 1$ corresponds to the Cauchy distribution, $P_1(x) = \frac{c}{\pi (c^2+x^2)}$. In general, all L\'evy stable distributions $P_{\mu}(x)$ with $\mu<2$ follow the
power-law asymptotic behavior 
\begin{eqnarray}\label{heavy tailed distribution}
\lim_{x\rightarrow \infty} P_{\mu}(x) \equiv \frac{c^{\mu} \Gamma (1+\mu) \sin \pi \mu/2}{x^{1+\mu}}~,
\end{eqnarray}
such that the variance is infinite and the fractional moments $\langle |x|^q \rangle$ are finite only for $q < \mu$ \cite{Jacobs}. Note that the
``tail'' of L\'evy densities are much wider than those of the Gaussian distribution \cite{Jacobs}.

In order to characterize the probability distribution of the random fluctuations $h(x,t)$, we start from Eq. (\ref{fractional Fokker-Planck equation}), and write the Langevin equation governing the dynamics of the Fourier components of the random field $h(k,t)=\int_{-\infty}^\infty e^{-ikx}h(x,t)dx$ which evolve according to
\begin{eqnarray}\label{fourier FEW langevin}
\dot{h}(k,t) = -D|k|^{z} h(k,t) +\eta_{\mu}(k,t)~,
\end{eqnarray}
where $\eta_{\mu}(k,t)$ is power-law uncorrelated noise (in Fourier space) with the symmetric heavy-tailed distribution given in Eq. (\ref{heavy tailed distribution}) at a particular time instant. For a fixed value of the Fourier mode $k$, it is convenient to discuss Eq. (\ref{fourier FEW langevin}) in terms of the Langevin approach to L\'evy flights in harmonic potential, such that
\begin{eqnarray}
\dot{\mathcal{Q}} = f(\mathcal{Q}) +\tilde{\eta}_{\mu}(t)~,
\end{eqnarray}
where $\mathcal{Q}$ is the stochastic random process, $f(\mathcal{Q}) = -D_k \mathcal{Q}$ is the harmonic force and $\tilde{\eta}_{\mu}(t)$ is the white L\'evy stable noise. In many situations, one is interested in the probability distributions  $\mathcal{P}(\mathcal{Q},t)\equiv \mathcal{P}(h,k,t)$ at some
given time $t$. It can be shown that the kinetic equation for the probability distribution of the stochastic process $\mathcal{Q}$ has the form 
\begin{eqnarray}\label{fractional Fokker-Planck equation}
\frac{\partial}{\partial t} \mathcal{P}(\mathcal{Q},t) = \left( -\frac{\partial}{\partial \mathcal{Q}} f(\mathcal{Q}) + K_{\mu}\frac{\partial^\mu}{\partial |\mathcal{Q}|^\mu} \right) \mathcal{P}(\mathcal{Q},t)~,
\end{eqnarray}
which is called space-fractional Fokker-Planck equation \cite{Klages}. Stationary solution of the fractional diffusion equation Eq. (\ref{fractional Fokker-Planck equation}) can be read from the asymptotic time independent solutions $\mathcal{P}_{st} (\mathcal{Q}) = \lim_{t\rightarrow \infty} \mathcal{P} (\mathcal{Q}, t)$. By analogy with the Ref. \cite{Chechkin2002}, the asymptotic limit ($\mathcal{Q}\rightarrow \infty$) of the stationary solution to the fractional Fokker-Planck equation (\ref{fractional Fokker-Planck equation}) reads
\begin{eqnarray}\label{heavy tailed distribution Q}
\mathcal{P}_{st} (\mathcal{Q}) =  \frac{C_\mu}{D_k} \frac{1}{|\mathcal{Q}|^{1+\mu}}~,
\end{eqnarray}
where $C_\mu = \frac{K_\mu \sin\pi\mu/2 \Gamma(\mu)}{\pi}$ and $D_k = D|k|^z$. Thus, the stationary solution $\mathcal{P}_{st} (\mathcal{Q})\equiv \mathcal{P}_{st} (h,k)$ is a symmetric L\'evy distribution, see Eqs. (\ref{heavy tailed distribution}) and (\ref{heavy tailed distribution Q}). Thus, we expect that the stationary probability distribution of the fluctuating field (with dynamics governed by a linear Langevin equation \ref{fractional edwards wilkinson}) is given by 
\begin{eqnarray}\label{power law saturation pdf}
\mathcal{P}_{st} (h) \propto 1/h^{1+\mu}.
\end{eqnarray}
It turns out, that the random field $h(x,t)$ given by the solution to Eq. (\ref{fractional edwards wilkinson}) is
clearly a correlated $\mu$-stable process (for $0<\mu<2$).

\section{Numerical Results}\label{section 4}

\begin{figure}[t]
\begin{center}

\includegraphics[width=9cm,clip]{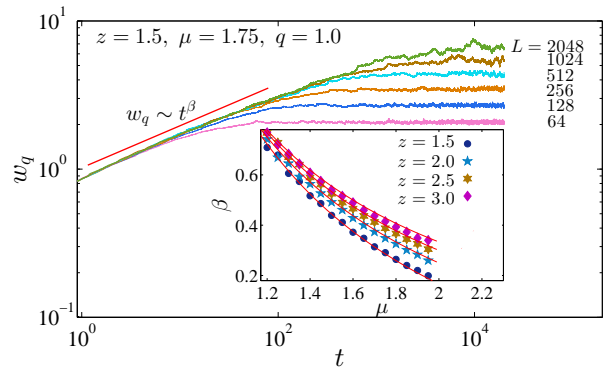}
\end{center}
\caption{(Color online) The $q$th order global width $w_q(t,L)$ with respects to time $t$ ($q=2/\mu$ for each case). For early times $t\ll L^z$,  the $q$th order order global width scaled according to $w_q(t)\sim t^{\beta}$. Inset shows the scaling exponent (growth exponent) $\beta$ for different values of the fractional order $z$ and the stable parameter $\mu$. The solid red lines show the theoretical prediction $\beta= \frac{z+1}{z\mu}-\frac{1}{z}$.}
\label{fig1}
\end{figure}
\begin{figure}[t]
\begin{center}

\includegraphics[width=9cm,clip]{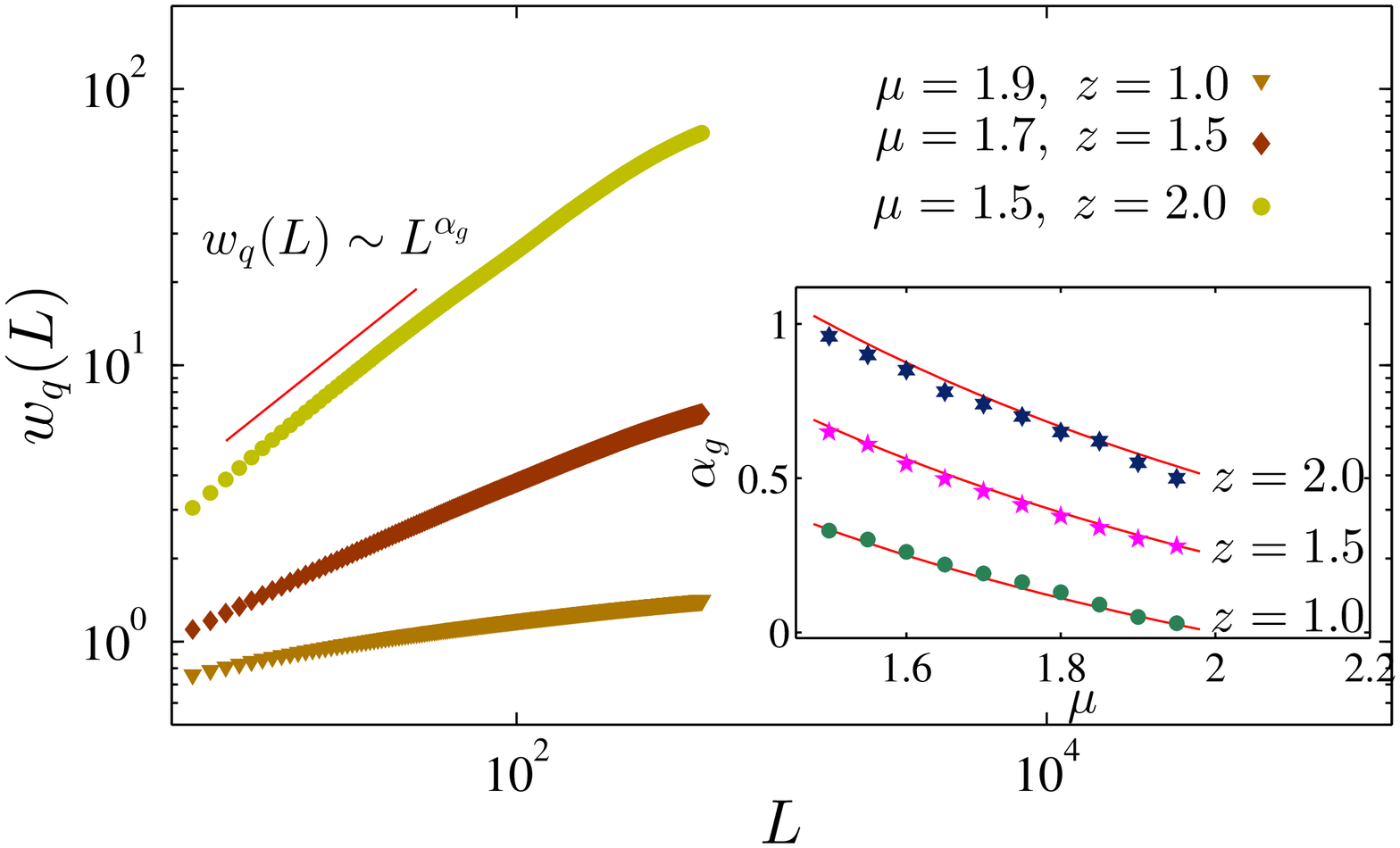}
\end{center}
\caption{(Color online) The $q$th order order global width $w_q(L)$ in the steady-state regime ($q=2/\mu$ for each case). Solid red line represents the scaling behavior $w_q(L)\sim L^{\alpha_g}$. Inset shows the scaling exponent (global roughness) $\alpha_g$ for different values of the fractional order $z$ and the stable parameter $\mu$. The solid red lines show the theoretical prediction $\alpha= \frac{z+1}{\mu}-1$.}
\label{fig2}
\end{figure} 
In order to check the above predictions, this part is concerned with the numerical simulations of non-local diffusion equation in the presence of the $\mu$-stable L\'evy noise for different fractional order $z$ and stable exponent $\mu$. To this end, we transform the Langevin equation (\ref{fractional edwards wilkinson}) to discrete time and space. Consider the one-dimensional lattice of size $L$ and grid size $\Delta x$, with periodic boundary condition. We define $x = m\Delta x$ with $m=1,\dots,L$ and $t=n\Delta t$ with $n=0,1,\dots,N$, where $\Delta t$ is the time step.

The discretization of the Riesz-Feller derivative, can be efficiently generated by the matrix transform algorithm \cite{ilic2005numerical,ilic2006numerical,nezhadhaghighi2014}, which is,

\begin{eqnarray}\label{discretGEM}
\left. \frac{\partial ^z}{\partial |x|^z}h(x,t) \right|_{(x=i\Delta x,t=n\Delta t)} \equiv \frac{1}{(\Delta x)^z}\sum_{j= 1}^{L} \mathbb{K}_{i,j}h_j^n~, 
\end{eqnarray} 
 where the elements of the matrix $\mathbb{K}$, representing the discretized fractional operator, can be expressed as,
\begin{eqnarray} \label{HOLR} 
\mathbb{K}_{l,m}=\frac{\Gamma(-\frac{z}{2}+n)\Gamma(z +1)}{\pi \Gamma(1+\frac{z}{2}+n)} \sin(\frac{z}{2}\pi),
\end{eqnarray}
where $n=\vert l-m\vert$, and fractional order $z\geq 1$. If $z/2$ is an integer, then 
$\mathbb{K}(n) = (-1)^{z-n+1}C_{z,\frac{z}{2}+n}$ for $n\leq z/2$ and $\mathbb{K}(n)=0$ for $n> z/2$, where $C_{z,\frac{z}{2}+n}$ are binomial coefficients \cite{zoia2007fractional}.

If the differential equation (\ref{fractional edwards wilkinson}) is discretized in time using an explicit (Euler)
method, besides Eq. (\ref{HOLR}), we obtain a finite difference approximations to the fractional diffusion
equation, that is,

\begin{eqnarray}\label{discrete FDE}
h_i^{n+1} = h_i^n +  \frac{\Delta t}{(\Delta x)^z} \sum_{j= 1}^{L} \mathbb{K}_{i,j}h_j^n + \Delta t^{1/\mu} \Delta x^{1/\mu -1}\zeta~,
\end{eqnarray}
where $h_i^n$ approximates the random fluctuation $h(x,t)$ in terms of discrete space and time variables $x=i\Delta x$ and $t= n\Delta t$. In order to numerically generate the  uncorrelated random noise $\eta(x,t)$ with heavy-tailed distribution, using the approximation discussed in Appendix. \ref{Appendix 1},  we used $\eta(x,t) \sim (\Delta x \Delta t)^{1/\mu -1} \zeta$ where $\zeta$ is independent identically random variable with distribution in the family (\ref{power law PDF}). 

Although the probability densities for almost all the $\mu$-stable processes have no known closed
form, there is a simple and efficient method for numerically generating uncorrelated random numbers $\zeta$ with the symmetric L\'evy distribution \cite{Jacobs}. These are given by 
\begin{eqnarray}
\zeta = \frac{\sin(\mu V)}{\cos(V) ^{1/\mu}} \left[\frac{\cos((1-\mu)V)}{ W} \right] ^ {(1-\mu)/\mu},
\end{eqnarray}
where $V$ has a uniform density on the interval $(-\pi/2,\pi/2)$, and $W$ has the standard exponential distribution with mean $1$ \cite{weron}.

Here, for sake of simplicity, the lattice constant $\Delta x$ has been set equal to one. The length of the time step, $\Delta t$, must be small enough (we choose $\Delta t = 0.01$) to ensure the stability in Eq. (\ref{discrete FDE}). In order to determine scaling characteristics of the scalar field $h(x,t)$, we have simulated this model on a lattice of size $L \in {64,128,256,512,1024,2048}$, with flat initial condition ($h(x,0)=0$). All numerical measurements are made using an ensemble of $2\times 10^4$ realizations. As already mentioned earlier, in order
to minimize finite-size effects, we impose periodic boundary conditions in such a
way that for each elements of the matrix kernel $\mathbb{K}$, the relation $\mathbb{K}(n) = \mathbb{K}(n-L)$ occurs. 

\begin{figure}[t]
\begin{center}

\includegraphics[width=9cm,clip]{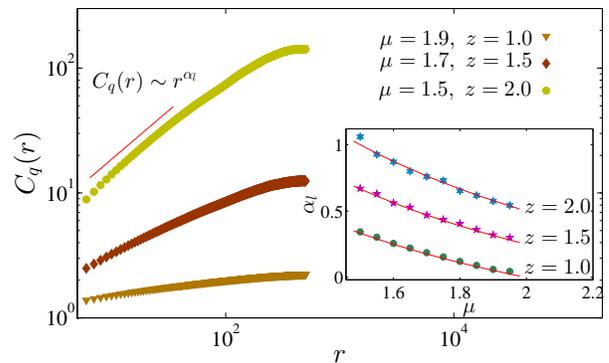}
\end{center}
\caption{(Color online) 
The $q$th order correlation functions $C_q(r)$ in the saturated regime ( $q=2/\mu$ for each case). Solid red line represents the universal scaling behavior $C_q(r)\sim r^{\alpha_l}$. Inset shows the scaling exponent (local roughness) $\alpha_l$ for different values of the fractional order $z$ and the stable parameter $\mu$. The solid red lines show the theoretical prediction $\alpha= \frac{z+1}{\mu}-1$.}
\label{fig3}
\end{figure}

\subsection{Scaling exponents and self-affinity}
We will now consider in detail the procedure for extracting the scaling exponents, $\beta$ and $\alpha_g$, from the $q$th order global width Eq. (\ref{global roughness scaling}) and, $\alpha_l$ from the $q$th order two-point function Eq. (\ref{local roughness scaling}). 

From the scaling properties the $q$th order global width, Eq. (\ref{global roughness scaling}), we can
measure growth exponent $\beta$. A plot of the global width $w_q$, with respect to time $t$, is given in Fig. (\ref{fig1}) for various system size $L$. Through the relation $w_q\sim t^{\beta}$ for early times $t\ll L^\nu$, we obtain the exponent governing the rate of growth of the interface width. Since the log-log plot of $w_q$ is a linear function of time, we calculate $\beta$ as a function of stable parameter $\mu$  and fractional order $z$ in Fig. (\ref{fig1}). Our numerical results agree well with the theoretical prediction $\beta= \frac{z+1}{z\mu}-\frac{1}{z}$.

To determine the global roughness exponent $\alpha_g$ describing the saturation of the $q$th order global width, we use the relation $w_q \sim L^{\alpha_g}$ in the steady-state regime $t\gg L^z$. In Fig. (\ref{fig2}) we show the scaling behavior of $q$th order global width $w_q$ with respect to the system size $L$ in the $\log-\log$ scale. In Fig. (\ref{fig2}) we show $\alpha_g$ versus $\mu$ for different values of the fractional order $z$. The numerical simulation results were found to agree well with
the analytical prediction $\alpha= \frac{z+1}{\mu}-1$. 

To check the validity of the scaling formulas derived in Sec. \ref{section 2}, and to ensure the self-affinity of the correlated random fluctuations with heavy-tailed distributions, we have numerically evaluated the $q$th order two-point function Eq. (\ref{local roughness scaling}). In the saturation regime, In Fig. (\ref{fig3}) we show our numerical results for the correlation function $C_q$ scaled according to $C_q(r)\sim r^{\alpha_l}$ where $\alpha_l$ is the local roughness exponent. 
As shown in Fig. (\ref{fig3}), we find excellent agreement between
measured values of $\alpha_l$ with theoretical predictions $\alpha= \frac{z+1}{\mu}-1$. The self-affine scaling property of the heavy-tailed random fluctuation $h(x,t)$ obtained from $\alpha_l=\alpha$ for all $q$s.

\subsection{Heavy-tailed distribution function}
\begin{figure}[t]
\begin{center}

\includegraphics[width=9cm,clip]{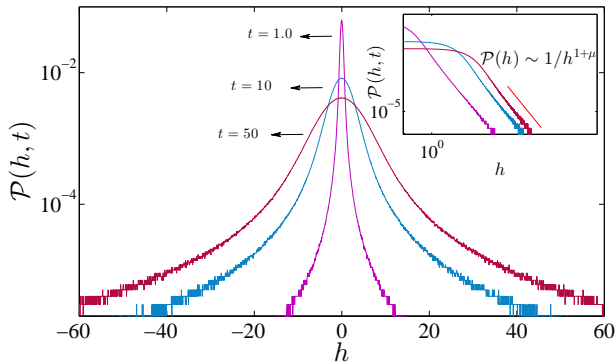}
\end{center}
\caption{(Color online) The probability distribution function $\mathcal{P}\lbrace h,t \rbrace$ for the random fluctuation $h$ in semi-log scale. The heavy-tailed random fluctuation $h(x,t)$ is the solution to the Eq. (\ref{fractional edwards wilkinson}) with $z=1.5$ and $\mu = 1.55$. (Inset) The same figure in $\log$-$\log$ scale shows this distribution is not Gaussian and it is distributed according to a power-law form $\mathcal{P}\lbrace h \rbrace \sim 1/h^{1+\mu}$.}
\label{fig4}
\end{figure}
We now present the special properties of distribution function of the random fluctuation $h(x,t)$. As we saw earlier, probability distribution $\mathcal{P}(h,t,L)$ is essential for understanding the behavior of the dynamics of the process Eq. (\ref{fractional edwards wilkinson}). Specifically, using Fourier component $\mathcal{Q} \equiv h(k,t)$, the time evolution of the corresponding probability distribution $\mathcal{P}(\mathcal{Q},t)$ is governed by the fractional Fokker-Planck equation (\ref{fractional Fokker-Planck equation}). Therefore, we expect the stationary solution to the fractional Langevin equation (\ref{fractional edwards wilkinson}) in the presence of the $\mu$-stable L\'evy noise, to be distributed according to a power-law $\mathcal{P}(h)\sim {1/h^{1+\mu}}$. Fig. (\ref{fig4}) shows $\mathcal{P}(h,t)$ as functions of $h$ for several times. These distributions are not Gaussian as indicated by the heavy-tails of the PDFs. These tails can be well approximated by power-law scaling behavior (\ref{power law saturation pdf}). So, in the saturation regime, $\mathcal{P}_{st}(h)$ increases linearly on a log-log plot, with the slope equal to $1+\mu^\prime$. We observe that the scaling factor $\mu^\prime$ behaves equal to the stable L\'evy distribution index $\mu$.

\subsection{Diffusion entropy analysis}

\begin{figure}[t]
\begin{center}

\includegraphics[width=9cm,clip]{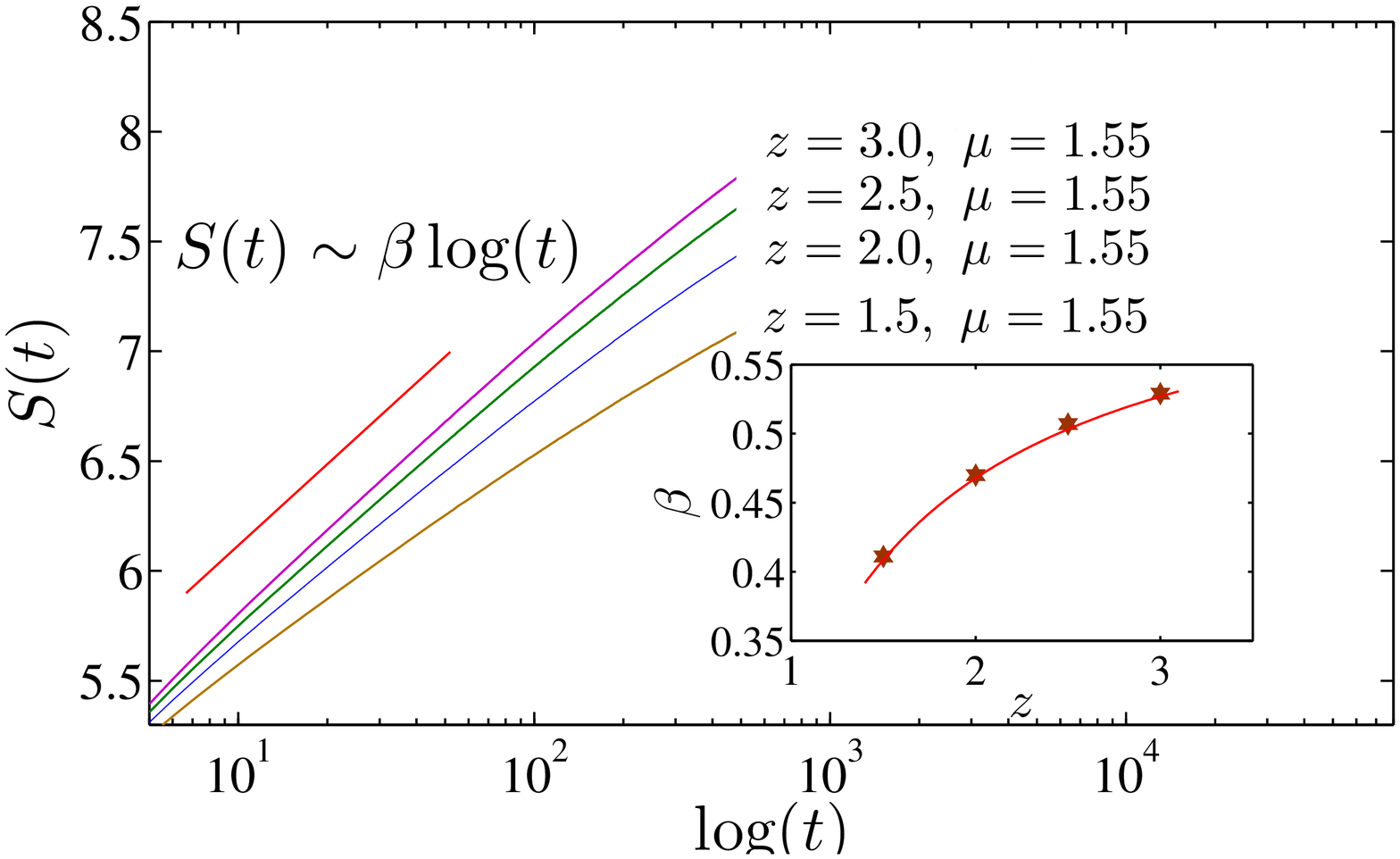}
\end{center}
\caption{(Color online) Time evolution of diffusion entropy $S(t)$ for a numerical solution to Eq. (\ref{fractional edwards wilkinson}). When the probability distribution is a power law
in time (see Eq. (\ref{scaling of prob h})), the diffusion entropy increases linearly on a log-linear plot: $S(t) \sim \beta \log (t)$. Inset shows the scaling exponent (growth exponent) $\beta$ for different values of the fractional order $z$. The solid red lines show the theoretical prediction $\beta= \frac{z+1}{z\mu}-\frac{1}{z}$ (with $\mu = 1.55$).}
\label{fig5}
\end{figure}
As a final characterization of the scaling behavior
for the probability distribution $\mathcal{P}(h,t,L)$, we shall discuss here the special form of PDF: $ \mathcal{P}(h,t,L) = \frac{1}{t^\beta L^\alpha }\mathcal{F}\left( \frac{h}{t^\beta},\frac{h}{L^\alpha} \right)$. As has been already shown, the early time dynamics of the probability distribution function, can be written in the scaling form of $\mathcal{P}_{er}(y,t) = \frac{1}{t^\beta }\mathcal{G}_1\left( \frac{y}{t^\beta} \right)$ (see Eq. \ref{scaling of prob h: early time}). Very interestingly, when the scaling function is a power law in time, the diffusion entropy $S(t) = -\int dx p(x,t)\log p(x,t)$ provides an estimation for the scaling exponent $\beta$ which is exactly the same as the growth exponent. Figure (\ref{fig5}) illustrates diffusion entropy analysis of the correlated random fluctuations with heavy-tailed distribution which exhibit the log-linear scaling property $S(t)\propto \beta \log t$. It is an interesting fact to stress here that the scaling exponent $\beta$ closely follows the theoretical prediction
given by $\beta = \frac{z+1}{z\mu}-\frac{1}{z}$. In Fig. (\ref{fig5}) the prefactor $\beta$ is plotted as a function of fractional order $z$. 

\section{Conclusion}\label{section 5}
To summarize, we have studied the scaling properties of the random fluctuations with power-law distribution functions. By considering the non-local fractional operator and the $\mu$-stable L\'evy noise, in a rather simple linear langevin equation (\ref{fractional edwards wilkinson}), we have been able to characterize the statistical properties of non-Gaussian heavy-tailed random fluctuations. First, through scaling arguments we have shown that the solution to the fractional diffusion equation (\ref{fractional edwards wilkinson}) obeys a simple scaling law, with a scaling exponent $\alpha$. We expect that the $q$-th order global moment and the $q$-th order correlation functions obey the power-law scaling. We derive analytic expressions for the local scaling exponents. We have also shown how the roughness exponent $\alpha$ and growth exponent $\beta$ can be analytically derived. We performed numerical simulations and measured these scaling exponents. Our numerical results are
in perfect agreement with the analytical calculations, which supporting the numerical scheme developed herein.
We also analyzed the scaling properties of the probability distribution functions and demonstrated the heavy-tailed distributions of the process by investigating the fractional Fokker-Planck equation for non-Gaussian heavy-tailed random fluctuations. We have shown that the probability distribution function for the solution to Eq. (\ref{fractional edwards wilkinson}), even if the process is correlated in space and time, scales like $\mu$-stable L\'evy processes.  Based on the diffusion entropy analysis, we have studied the special scaling properties of the probability distribution functions, and we numerically calculated the growth scaling exponent $\beta$. The excellent agreement between the theoretical predictions and the numerical simulations provides a validation of our simulations. 
Finally, our study gives useful insights into the correlated random fluctuations with power-law PDFs, which makes it possible to establish the genuine nature of the related Physical phenomenons. This is
left as a subject for further studies.

\appendix
\section{Scaling properties of $\mu$-stable L\'evy noise}\label{Appendix 1}

Here, we discuss the scaling behavior of the symmetrical L\'evy stable process $\eta_\mu (x,t) = \lambda ^{-\gamma} \eta_\mu (\lambda x,\lambda^\nu t)$ under a scale transformation $x\rightarrow \lambda x$ and $t\rightarrow \lambda^\nu t$ \cite{lam1993exact}. First consider a small region $\Omega=\Delta x \Delta t$ in ($x-t$) space, so we need to perform integration $I = \int_{\Omega} \eta_\mu (x,t) dx dt$. We can easily approximate it using the Monte Carlo method that numerically computes a definite integral. Thus, we could use the following procedure: (\textit{a}) consider $N\sim \Delta x \Delta t$ random points on $\Omega$ with uniform distribution, (\textit{b}) generate $N$ random numbers ${\xi_i}$ from power-law distribution $P(x)\sim 1/|x|^{1+\mu}$, (\textit{c}) the integral $I$ can be approximated by 
\begin{eqnarray}
\int_{\Omega} \eta_\mu (x,t) dx dt= \sum_{i=1}^N \xi_i \sim N^{1/\mu} \xi. 
\end{eqnarray}
where the latter approximation is obtained using the properties of the stable laws and the generalization of the central limit theorem \cite{stochastic book,gardiner}. 
This relation should be invariant under the scale transformation:
 \begin{eqnarray}
\lambda^{1+\nu}\int_{\Omega} \eta_\mu (\lambda x,\lambda ^z t) dx dt \equiv \lambda^{(1+\nu)/\mu} N^{1/\mu} \xi. 
\end{eqnarray}
which implies $\gamma = (1/\mu -1)(1+\nu)$ \cite{lam1993exact}.


\end{document}